\documentstyle[12pt,aaspp4,psfig]{article}
\begin{document}

\def\Sbar{{\bar{S}}}
\def\Fbar{{\bar{F}}}
\def\Pne{P_{\delta n_e}}
\def\smperp{{SM_{\perp}}}
\def\smun{{kpc \,\, m^{-20/3}}}
\def\cnsq{{C_n^2}}
\def\nbar{{\overline{N}}}
\def\dne{{\delta n_e}}
\def\FWHM{{\theta_{FWHM}}}
\def\nelos{{\langle n_e \rangle_{l.o.s.}}}

\def\rsun{{{R}_{\odot}}}
\def\vsun{{{\bf V}_{\odot}}}
\def\msun{{M_{\odot}}}
\def\eps02{{{\epsilon_{-2}}}}
\def\pspin{{{P_{spin}}}}
\def\porb{{P_{orb}}}
\def\Omegaspin{{\Omega_{spin}}}
\def\Omegaorb{{\Omega_{orb}}}
\def\Omegaspindot{{\dot{\Omega}_{spin}}}
\def\Omegaorbdot{{\dot{\Omega}_{orb}}}
 
\def\kms{{km s$^{-1}$ }}
 
\def\mathbf{{\boldmath}}

\def\mura{{\mu_{\rm RA} }}
\def\mudec{{\mu_{\rm DEC} }}
\def\errmura{{\sigma_{\mu_{RA}} }}
\def\errmudec{{\sigma_{\mu_{DEC}} }}

\def\sign{{\sigma_{N} }}
\def\signprime{{\sigma_{N}^{\prime} }}

\def\muw{{{\mu}_W}}
\def\errmuw{{\sigma_{\mu_{W}} }}

\def\dhat{{\hat D}}
\def\taus{{\tau_S}}
\def\vperp{{V_{\perp}}}
\def\vr{{V_r^{(P)}}}
\def\vri{{V_r}}
\def\vrd{{V_r}^{(DGR)}}
\def\vp{{ \mbox{\boldmath $V$}}$_{\perp}^{(P)}$}
\def\vpmath{{ \mbox{\boldmath V}}_{\perp}^{(P)}}
\def\cl{{\cal L}}
\def\cd{{\cal D}}
\def\zb{{z_0}}
\def\vkick{{V_{\rm kick}}}

\def\vperpvec{{ \mbox{\boldmath $V_{\perp}$} }}
\def\dmuvec{{ \mbox{\boldmath $\delta\mu$} }}
\def\muvec{{ \mbox{\boldmath $\mu$} }}
\def\data{{\tilde \muvec}}
\def\dataw{{\tilde{\mu}_W}}

\def\errmu{{\sigma_{\muvec} }}

\def\thetavec{{ \mbox{\boldmath $\theta$} }}

\def\zhat{{\bf \hat z}}
\def\nhat{{\bf \hat n}}

\def\be{\begin{eqnarray}}
\def\ee{\end{eqnarray}}

\def\ld{{\ell_d}}
\def\inner{{\ell_1}}
\def\outer{{\ell_0}}
\def\lf{{\ell_F}}
\def\dnud{{\Delta\nu_{\rm d}}}

\def\dnu{{\delta\nu}}
\def\dtd{{\Delta t_{\rm d}}}
\def\dt{{\delta t}}
\def\dte{{\delta t_e}}

\def\tb{{{\rm T}\times{\rm B}}}
\def\phirms{{\phi_{\rm rms}}}

\def\evec{{ {\bf E}}}
\def\sigmaF{{\sigma_F}}
\def\Nprime{N^{\prime}}
\def\Nbarprime{\nbar^{\prime}}
\def\dNpdf{{f_{\Delta {\rm N}}}}
\def\dN{{\Delta {\rm N}}}
\def\dF{{\Delta {\rm F}}}
\def\niss{{{N_{\rm ISS}}}}

\title{A VLA Search for the Geminga Pulsar: A Bayesian Limit on a Scintillating 
Source}
\author{M. A. McLaughlin\altaffilmark{1}, J. M. Cordes\altaffilmark{1}, T. H. 
Hankins\altaffilmark{2}, \& D. A. Moffett\altaffilmark{3}}
\altaffiltext{1}{Astronomy Department, Cornell University, Ithaca, NY 14853}
\altaffiltext{2}{Physics Department, New Mexico Institute of Mining and 
Technology, Socorro, NM 87801}
\altaffiltext{3}{School of Physics, University of Tasmania, Hobart TAS 7001, 
Austrailia}
\begin{abstract}

We derive an upper limit of 3 mJy (95\% confidence) for the flux density at 317 
MHz of the
Geminga pulsar (J0633+1746). Our results are based on 7 hours of fast-sampled 
VLA data,
which we averaged synchronously with the pulse period using a period
model based on CGRO/EGRET gamma-ray data. Our limit accounts for the fact that 
this  
pulsar is most likely subject to interstellar scintillations on a timescale much 
shorter
than our observing span. Our Bayesian method is quite general and can be applied 
to
calculate the fluxes of other scintillated sources. We also present a Bayesian technique for
calculating the flux in a pulsed signal of unknown width and phase.  
 Comparing our upper limit of 3 mJy with the quoted flux density of
Geminga at 102 MHz, we calculate a lower limit to its spectral index of $\alpha 
\approx 
2.7$ ($F(\nu) \propto \nu^{-\alpha}$). 
We discuss some possible reasons for Geminga's weakness at
radio wavelengths, and the likelihood that many of the unidentified EGRET 
sources
are also radio-quiet or radio-weak Geminga-like pulsars.

\end{abstract}

\section{Introduction}

Geminga has long been known as a strong X-ray and gamma-ray continuum source
(\cite{fic75}) and, in 1987, was identified with an optical star (\cite{big87}). 
In 1992,
Geminga was detected at X-ray (\cite{hal92}) and gamma-ray (\cite{ber92}) 
energies as
a 237-ms pulsar with a period derivative of 
$10.95\times10^{-15}$ s s$^{-1}$, indicating a characteristic age of 
$3.4\times10^{5}$ yr
and a magnetic field of $1.6\times10^{12}$ Gauss. A proper motion of $170 \pm 6$ 
mas yr$^{-1}$
and
a parallax of $6.36 \pm 1.7$ mas have been measured
for Geminga's optical counterpart (\cite{car96}). From these measurements, a 
distance
of 157$^{+59}_{-34}$ pc and a transverse velocity of $140 \pm 15$  km s$^{-1}$ 
have been
calculated.

Since its discovery as a gamma-ray source in 1975, many observers have 
unsuccessfully
attempted radio
detection of Geminga both as a continuum source (\cite{spel84}) and
as a pulsar (\cite{seir92}). Recently, however, three independent groups 
(\cite{mal97};
\cite{kuz97}; \cite{shit97a}) have reported successful detections of pulsed 
radio emission
from Geminga. All 3 detections were made at the 
Pushchino Radio Astronomy Observatory at a radio frequency of 102 MHz, a lower 
frequency
than those used for previous searches.
All three groups calculate a dispersion measure (DM) of roughly 3 pc cm$^{-3}$. 
The measured fluxes vary greatly, perhaps due to interstellar scintillations,
and range from 5-500 mJy. Reported pulse widths range from 10 to 120 ms.
Recently, Shitov \& Malofeev (1997)\nocite{shit97b} reported a 41 MHz detection
of Geminga with mean flux density
300 mJy. Comparing this result with Malofeev and Malov's quoted 102 MHz mean flux of 60 
mJy, we
may estimate a spectral index of $\alpha \approx 1.8$, where $F(\nu) \propto 
\nu^{-\alpha}$.
Comparing Malofeev and Malov's quoted mean flux of 60 mJy with the previous 1 
mJy upper
limit at 1.4 GHz (\cite{seir92}), we calculate a lower bound on Geminga's  
spectral index of $\alpha \approx 1.6$.

Since Geminga is the {\it only} known gamma-ray pulsar which is not also a 
strong radio
source, confirming its existence as a radio pulsar and determining the reasons 
for
its radio weakness are essential to understanding the 
relationship between pulsar radio and gamma-ray emission. Geminga is also 
important as it
may be a prototype for a class of radio-weak or radio-quiet high-energy pulsars 
which
could account for some of the 80 EGRET sources not yet identified with known 
pulsars or
active galactic nuclei.

We have therefore revisited
the search for Geminga as a radio pulsar with a long, low frequency, 
fast-sampled 
VLA observation. In this paper, we present our results, introducing  Bayesian 
methods for
determining
the pulsed flux in a signal of unknown width and phase and
for estimating the intrinsic flux density of a source, accounting for 
interstellar
scintillations.

\section{Data}

We observed Geminga (PSR J0633+1746) on 2 February 1998 with the D-configured 
VLA 
in phased array mode. The data spanned a bandwidth of 25 MHz, centered on
317.5 MHz, with 14 (not necessarily consecutive) 1-MHz channels
(situated to avoid known interference 
frequencies).
 Full polarization data were recorded with the High Time Resolution
Processor (\cite{mof97}) using a sampling frequency of 1152 Hz. We obtained
7 one-hour observations (scans) of Geminga which spanned a total time of 9 hours.
Two other known pulsars (B0329+54 and B0950+08) were observed
for 10 minutes each with the same setup as that used for Geminga.
 Each scan was started on a 10 second tick to ensure accurate pulse
phase referencing between scans. This 10-second tick was tied to the VLA's hydrogen
maser, which is compared to Universal Time through the GPS.
 The array was phased, with rms
phase errors less than 1$^{\circ}$, between consecutive scans. We observed a 
strong VLA calibrator source (B0813+482) with known flux density ($S$ = 45 Jy)
to calculate the gain in each of the 56 channels (14
frequencies and 4 polarizations).
As we expect only a 7-mJy uncertainty in our flux density calibration due to 
radiometer noise,
our main source of uncertainty is the intrinsic source flux variation. By 
comparing the
VLA flux measurement with those from the Westerbork and Texas surveys 
(\cite{ren97};
\cite{doug96}), we estimate a maximum intrinsic flux density variation of 10\%.
We therefore expect errors in our final reported flux densities to be less than 
10\%.

\section{Analysis}

For each of the 7 Geminga data sets, we summed polarizations to calculate the 
total intensity
and dedispersed all 14 frequency
channels using a DM
of 3.2 pc cm$^{-3}$, the midpoint of the DM range predicted by the TC 
(\cite{tay93}) 
model for Galactic electron density given Geminga's measured distance
and direction.
Our estimated DM is consistent with those reported for the 102 MHz Geminga
detections. We note than even a 1 pc cm$^{-3}$ error
in our assumed DM would produce only 7 ms of pulse smearing, negligible for 
Geminga's
period of 237 ms. 

 Each one-hour dedispersed time series was folded using the gamma-ray
ephemeris of Mattox et al. (1998) (P = $0.23709574610$ s,  $\dot{\rm P} =
10.974012\times10^{-15}$ s s$^{-1}$, epoch = 2446600). We used a single period to 
fold each 1
hour data set, but updated the period between scans, using the gamma-ray 
ephemeris and TEMPO 
(\cite{tay89}). We note that using a single period for each scan results in only 
0.2 ms of pulse smearing.
Table 1 lists the modified Julian
date of the midpoint of each
scan, the corresponding period, and the total length of the scan.
The first 6 scans are 1 hr in length, while the last is 50 minutes, yielding
a total integration time of 6 hr 50 min. 

\begin{deluxetable}{cccc}
\tablewidth{4.0in}
\tablenum{1}
\tablehead{
\colhead{Scan number} & \colhead{MJD} & \colhead{Period (s)} & \colhead{Time
(s)}}
\startdata
1 & 50847.08182 & 0.237113398987 & 3600 \nl
2 & 50847.13715 & 0.237113501528 & 3600 \nl
3 & 50847.18646 & 0.237113605070 & 3600 \nl
4 & 50847.23252 & 0.237113704454 & 3600 \nl
5 & 50847.27766 & 0.237113797450 & 3600 \nl
6 & 50847.34061 & 0.237113908108 & 3600 \nl
7 & 50847.38646 & 0.237113967595 & 3000 \nl
\enddata
\end{deluxetable}{}

Figure 1 shows the folded profiles for scans 1 through 7, in addition to the 
composite
profile formed by calculating the phase at the start of each scan, and shifting 
and
summing the profiles accordingly. We have verified that our phase-shifting 
algorithm
is correct by applying it to the observatory generated 19.2 Hz
waveguide switch signal. There is structure in some of the profiles of Figure 1. 
However,
no consistent pulse shape and/or phase is apparent. Furthermore, coherently 
summing all
profiles dramatically decreases the signal-to-noise level of any features. We 
have
also calculated Fourier transforms of all 7 dedispersed time
series. We have not detected any harmonics of Geminga's 237-ms period. 

Figure 2 shows the folded pulse profiles for our test pulsars, B0329+54 (DM = 
26.8 pc 
cm$^{-3}$) and B0950+08 (DM = 3.0 pc cm$^{-3}$).
These data were processed with identical
dedispersion and folding routines to the Geminga data. We have subtracted the 
off-pulse mean
to calculate $F$, the phase-averaged pulsed flux density.
For PSR B0329+54, $F$ = 1071 mJy, and for PSR B0950+08, $F$ = 8.9 mJy.
We note that we also detected both pulsars with high signal-to-noise in Fourier
transforms of  their time series.

Because Geminga may be too weak to be detected through its time-averaged flux, 
we searched for Crab-like `giant' pulses, or individual pulses with amplitudes 
much greater 
than the mean pulse amplitude (\cite{lund95}). We search for these
aperiodic, dispersed pulses by dedispersing each data set
over a range of trial DMs and, for each dedispersed time series, recording those 
pulses
with amplitudes above some signal-to-noise threshold.
We enchanced our sensitivity to broadened pulses by repeating this thresholding 
with different
levels
of time series smoothing. We found no evidence for isolated pulses from Geminga.
In Figure 3, we show the results of this analysis for Geminga,
PSR B0950+08, and PSR B0329+54, from which we detect a large number of 
individual pulses.

\subsection{Bayesian Pulsed Flux Estimation}

In Appendix A, we derive a Bayesian procedure for estimating the pulsed flux 
density
given a folded profile with a pulse of unknown amplitude, width, and phase. 
Assuming the
additive noise is Gaussian,
we are able to calculate a probability density function (PDF) for the
phase-averaged pulsed flux density {\it independent} of pulse width, pulse 
amplitude,
pulse phase, off-pulse mean, and off-pulse rms.
Because no prior assumptions are necessary to calculate an upper limit,
this method is an improvement over standard methods used to test for pulse shape 
uniformity
and to calculate upper limits.
The $\chi^{2}$ statistic, the method most commonly used, returns only the 
deviation of a
binned pulse profile from a uniform distribution. More sophisticated
tests, such as the Z$^{2}$ and H tests (\cite{dejager94}), developed to analyze 
sparse X-ray and gamma-ray 
profiles, are less dependent on binning and more sensitive
to a wide variety of pulse shapes than the $\chi^{2}$ statistic, but still 
yield only the
probability that a distribution differs from uniformity. To calculate an actual 
upper
limit, one must still assume a pulse width and phase. Furthermore, these methods 
do not provide a
simple mechanism for calculating a PDF for the pulsed flux density. 
In following papers, we will use our method to calculate more
accurate X-ray
and gamma-ray upper limits for known radio pulsars and to search for new high-energy
pulsars. The method can also be 
adapted to
a wide range of pulse shapes, in addition to the simple square pulse presented 
in Appendix A.

We apply the method of Appendix A to calculate PDFs for the pulsed flux density 
for the
folded Geminga pulse profiles shown in Figure 1. The resultant PDFs are also 
shown in Figure 1 and 
are discussed further in Section 3.2.2.

\subsection{Interstellar Scintillations of the Geminga Pulsar}

Since Geminga is a nearby (d $\approx$ 150 pc) pulsar, we expect its flux 
density to be strongly
modulated by diffractive interstellar scintillations (DISS). We therefore 
present a method
which allows us to calculate a PDF for a source's flux density given several 
measurements of
the scintillated flux density.  
We restrict our analysis to the strong scattering regime (a good assumption at 
low
radio frequencies), where $\phirms$,
the rms phase perturbation due to electron-density variations, is much greater 
than 1.
In this case, the amplitude fluctuations
are `saturated' because the electric field has been sufficiently randomized to
have complex Gaussian statistics.

We describe DISS through a characteristic timescale $\dtd$ and  characteristic
bandwidth $\dnud$. These quantities depend upon the scattering measure (SM), 
the integrated turbulence strength (assuming a power spectrum for electron 
density
irregularities) along the line of sight to the pulsar. In the strong scattering 
regime,
and assuming a uniform distribution
of scattering material and a Kolmogorov power spectrum,
$\dtd$ and $\dnud$ scale with SM, observing frequency, distance, and pulsar
velocity as (\cite{cordes91}; \cite{cordes98})
\be
\dtd = 10^{2.52}\nu^{1.2}{\rm SM}^{-0.6}V^{-1} \rm \hspace{0.1in} s,
\label{eq:scinttime}
\ee
\be
\dnud = 10^{-0.775}\nu^{4.4}{\rm SM}^{-1.2}D^{-1} \rm \hspace{0.1in} kHz
\label{eq:scintbw}
\ee
where $\nu$ is the radio frequency in GHz, SM is the scattering measure in kpc 
m$^{-20/3}$,
 $V$ is the pulsar's transverse velocity in km s$^{-1}$, and D is its distance in kpc.
The scaling is such that a distant source with high velocity observed at a low
frequency has a
{\it shorter} characteristic bandwidth and timescale than a nearby, low velocity 
source 
observed at high frequency. 

Once we have characterized the DISS properties of a source (through a 
measurement or an
estimation using Eqs.~\ref{eq:scinttime} and ~\ref{eq:scintbw}), we  
calculate the posterior PDF of the intrinsic source flux density given several
measurements of the
scintillated flux density using the method described in Appendix B.
The source's intrinsic flux density, $S$, is modulated by interstellar 
scintillations by a
factor $g$, so that the measured flux density, $F$, is
\be
F = gS + N,
\ee
where $N$, the additive noise, may include both radiometer noise and radio 
frequency interference. We note that
this additive noise model holds only for the case where the time-bandwidth 
($TB$)
product of the measurements (ie. integration time $T$ times bandwidth $B$)
is large, $TB \gg 1$. 
In the strong scattering regime, the scintillation time-bandwidth
product, $\dtd\dnud$, which is the characteristic size of a ``scintle'' in the 
$\nu-t$ plane,
may or may not be
much larger than $TB$ for the measurement. When the time-bandwidth product of 
the
measurement is {\it smaller} than that of one DISS `scintle,'
the PDF for $g$ is a one sided exponential:
\be
f_g(g) = e^{-g}, \hspace{0.1in} g \ge 0
\label{eq:fgexp}
\ee
with CDF
\be
F_g = P\{<g\} = 1 - e^{-g}.
\ee

When $ TB \gg\dtd\dnud$, the measurements average over many scintles,
increasing  the number of degrees of
freedom from 2 (for unquenched DISS) to 2$\niss$, where 
\be
\niss \approx
       \left ( 1 + 0.2\frac{B}{\dnud} \right )
        \left ( 1 + 0.2\frac{T}{\dtd} \right ).
\label{eq:niss}
\ee
Then, $g$ is a $\chi^{2}$ random variable with $2\niss$ degrees of 
freedom whose PDF is
\be
f_g(g, \niss) =
   \frac{(g\niss)^{\niss}}{g\Gamma\left(\niss\right )} e^{-g\niss} U(g),
\label{eq:fg}
\ee
where $U(g)$ is the Heaviside function  and $\Gamma$ is the gamma function.
For pulsars at large distances or those observed at low frequencies, the 
quantity $\niss$
will approach infinity and $f_g$ will tend toward a delta function, 
$\delta(g-1)$. Figure 4 illustrates the dependence of $f_g$ on $\niss$.

\subsubsection{Results for Known Pulsars}

For the known pulsars B0329+54 and B0950+08, $\dtd$ and $\dnud$ have been 
measured
(\cite{stine96}; \cite{phil92}). Applying the known frequency scaling of these 
quantities
(see Eqs.~\ref{eq:scinttime} \&~\ref{eq:scintbw}), we 
estimate $\dtd = 142$ s, $\dnud = 25$ kHz for B0329+54 and $\dtd = 2757$ s, 
$\dnud =
102$ MHz for B0950+08 at 317 MHz.
Given these values and our observing
bandwidth ($B = 25$ MHz) and time ($T = 600$ s),
we may calculate $\niss$ from Eq.~\ref{eq:niss}. For B0329+54, $\niss = 371$,
$TB \gg \dnud\dtd$, and $f_g$ will approach a delta function. B0950+08 is much
closer to Earth, so its time-bandwidth product
$\dnud\dtd > TB$ and $\niss = 1.095$. Although this pulsar is likely 
in the transition regime from
weak to strong scattering, $f_g$ will approximate an exponential and will have a 
similarly long tail.

Using the method of Appendix B, we calculate the PDFs for the intrinsic source 
fluxes of these
pulsars given by our measurements.
In Figure 5, we plot $f_S(S)$, where $S$ is the intrinsic phase-averaged flux density.
For B0329+54,
the PDF of the intrinsic source flux density is strongly peaked at the measured 
flux density,
and follows a
roughly Gaussian distribution away from the peak. For B0950+08, the PDF for the 
intrinsic
source flux density again peaks at the measured flux,
but is characterized by a broad exponential
tail. We can calculate confidence intervals for the flux densities of
these pulsars from their PDFs. For B0329+54, the 95\% confidence interval
is 968 mJy $< S <$ 1190 mJy, while for B0950+08, the 95\% confidence
interval is 5.4 mJy $< S <$ 880.1 mJy. We note that these pulsars'
predicted 317 MHz fluxes from the Princeton Pulsar Catalog (\cite{taylor93}) 400 
MHz fluxes and
measured spectral indices (\cite{lor95})
are 1890 mJy for PSR B0329+54 and 590 mJy for PSR B0950+08.
While the predicted flux for PSR B0329+54 lies outside of our confidence 
interval, 
this discrepancy is consistent with the $\approx 40\%$ modulations
due to refractive interstellar scintillation that have been measured for this
pulsar (\cite{stine96}).
Refractive ISS is probably also important in modulating the flux of PSR
B0950+08, although its predicted flux does fall within our confidence interval.

\subsubsection{Results for Geminga}

Because the scintillation bandwidth and timescale have not been measured for 
Geminga,
we use Eqs.~\ref{eq:scinttime} and ~\ref{eq:scintbw} to predict the
characteristic timescale and bandwidth given our observing frequency, Geminga's
measured distance, and the SM predicted from the TC model.
For Geminga, we estimate SM $\approx 10^{-4.4}$ kpc m$^{-20/3}$ and, from 
Eq.~\ref{eq:scintbw},
$\dnud \approx$ 1.5 MHz. Given Geminga's transverse velocity of $\vperp \approx 
140$ km s$^{-1}$,
Eq.~\ref{eq:scinttime} yields $\dtd \approx$ 275 s. Therefore, $TB \gg 
\dnud\dtd$,
$\niss \approx 16$, and we expect statistical independence
between measurements. Applying Eq.~\ref{eq:totalint} to the distributions
for $f_{F}(F)$ (see Figure 1), we calculate $f_S(S)$ for each of the 7
observations individually.
These PDFs are plotted in Figure 6. Using Eqs.~\ref{eq:likenorm} and 
~\ref{eq:totalint},
we may use all 7 Geminga observations to calculate a composite PDF, shown in 
Figure 7.
This PDF does peak at a non-zero value of $S$. However, as shown in
Figure 10 (see Appendix A), the PDFs of simulated noise often peak at non-zero 
values.
We note that the rms
noise levels in the Geminga pulse profiles are roughly equivalent to those of 
the simulated
profiles in this figure. In addition, we expect more structure in the Geminga 
profiles
due to the large amount of radio frequency interference at 317 MHz.
Furthermore, the marginalized PDFs for pulse width
and phase for each of the 7 Geminga scans do not peak at a consistent
value of these quantities, as expected for a real pulsar. We therefore treat our
result as an upper limit and calculate 
a 95\% confidence upper limit to Geminga's pulsed flux of 3.0 mJy.

The results so far have assumed that the pulse width is completely unknown.
We have also calculated upper limits assuming reasonable values for Geminga's 
pulse width.
If we assume Geminga's radio pulse has the same width (roughly 180$^{o}$) as its
gamma-ray pulse, we calculate a 95\% confidence upper limit to the pulsed flux 
density of
4.0 mJy.  However, except in the case of the Crab, the radio pulse widths of the 
known
EGRET pulsars are much narrower than the gamma-ray pulse widths. For this 
reason, we also
calculate an
upper limit for a pulse width of 25$^{o}$ (calculated assuming the $w \propto 
P^{-1/2}$
scaling of pulse
width $w$ with period $P$ given by Biggs (1990)) of 1.6 mJy.

We note that the distribution of material along the line of sight will most 
likely {\it not} be
uniform, and $\dtd$ and $\dnud$ may be somewhat different from those predicted 
by 
Eqs.~\ref{eq:scinttime} and ~\ref{eq:scintbw}.
We therefore explore what happens in the limiting
case where the scintillation timescale is {\it longer} than the 9-hr total 
timescale of our
observations.
For this case, we use Eq.~\ref{eq:likeid} to calculate the PDF for intrinsic 
source flux
density given all 7 observations, again integrating over the PDFs of Figure 1.
The resultant PDF for this limiting case, also plotted in Figure 7,
is much broader than that derived assuming statistically independent scintillations. 
The 95\% confidence upper limit on Geminga's pulsed
flux for this case is 13.9 mJy.
However, for the scintillation timescale to be as large as 9 hrs, the scattering 
measure
would have to
be 10$^{-4}$ times smaller than that estimated using the TC model.
Alternatively, a timescale this long could also be
explained by a thin screen located only 0.021 pc
from the observer (\cite{cordes98}). Obviously, both of these
scenarios are very improbable, and the true 95\% upper limit is likely very 
close to
3 mJy.

\section{Discussion and Conclusions}

Comparing our 3 mJy upper limit to Geminga's flux density at 317 MHz with the 
mean flux
density of 60 mJy reported by
Malofeev \& Malov at 102 MHz, we
calculate a {\it lower} limit on the spectral index of Geminga of $\alpha 
\approx 2.7$ ($F(\nu)
\propto \nu^{-\alpha}$),
 comparable to the spectral index of the Crab.  
If the spectral index of Geminga is indeed this
high, it would be extreme, as the Crab and Geminga pulsars have quite different ages,
spin periods, and spin-down rates. 

In Figure 8, we plot our upper limit along with previously measured upper limits and 
detections. Unfortunately, while there have been many radio searches for Geminga,
there are few published upper limits. This is partly due to Geminga's migrating gamma-ray
positional error box. For example,
Boriakoff et al. (1984) searched for radio pulsars in Geminga's error box,
determined from COS B data,
but did not cover
the current, well-determined position.
Of special interest is the recent non-detection of Geminga at both 35 and 327 MHz
by Ramachandran et al. (1998). 
While their quoted 327 MHz upper
limit of 0.3 mJy is much lower than ours, the rms sensitivities of our 7-hr VLA observation
and their 6-hr Ooty observation are very similiar. For consistency, we have plotted
the upper limit
obtained through applying our method to their data, taking into account scintillation
and the unknown width and phase of Geminga's pulse.

The reason for Geminga's radio weakness is still debatable.
One explanation is that some gamma-ray pulsars simply do not emit at radio
energies. Halpern \& Ruderman (1993) suggest that Geminga's radio
emission is quenched by the copious pair production in the inner magnetosphere. 
However,
the transient dips in Geminga's soft X-ray profile (\cite{hal97})
suggest that the processes responsible for supplying $e^{\pm}$ to the inner 
magnetosphere
may be variable. Because the radio silence will be occasionally broken as this 
$e^{\pm}$
quenching plasma clears away, we may expect Geminga to be a {\it transient} 
radio emitter.
If Geminga continues to be the only high-energy pulsar found to exhibit this 
transient
phenomenon, studying its X-ray properties may be useful for determining if and 
why
radio emission is suppressed.

A more likely explanation for Geminga's radio weakness is simply that there is
some misalignment between the gamma-ray and radio pulsar beams.
Geminga's pulse shape in gamma-rays (like that of the Crab and Vela pulsars) is 
broad and
double-peaked, suggesting an origin in the ``outer gaps'' of the pulsar
magnetosphere near the light cylinder. Because radio emission is expected to be
associated with the open field line region centered on the magnetic poles, it is 
possible
that our line-of-sight intersects the gamma-ray beam only. Romani \&
Yadigaroglu (1995), using the outer gap model for pulsar gamma-ray emission, 
have
modeled the orientation and size of the radio and gamma-ray pulsar beams given 
radio and gamma-ray data on known pulsars. They find that, because the gamma-ray 
beam
is much wider than the radio beam, 45\% of young pulsars will be detected {\it 
only} at high
energies, while only 19\% of young pulsars will be detected at both radio and 
gamma-ray
energies. Scaling from the 5 EGRET detections of radio pulsars, Romani \& 
Yadigaroglu
expect 12 nonradio
pulsars to be visible in gamma-rays at flux levels comparable to the radio 
selected
objects. This implies that {\it most} of the $\approx$ 20 unidentified Galactic 
EGRET sources
are young radio-quiet or radio-weak pulsars like Geminga. This is consistent 
with other
studies which find that the properties, such as flux variability (\cite{mcl96})
and spectral index (\cite{mer96}), of many unidentified sources are similar to
those of the known EGRET pulsars. It has also been shown (\cite{yad97}) that 
many
of the unidentified sources
are associated with regions of massive star formation and death, which are 
expected
to be breeding grounds for pulsars. 

\acknowledgments

We thank J. Mattox and K. Xilouris for helpful discussions.
M. A. McLaughlin acknowledges support from an NSF fellowship.
The work was also supported by the National Astronomy and Ionosphere Center, 
which is
operated by Cornell University under cooperative agreement with the National 
Science
Foundation (NSF). The National Radio Astronomy Observatory is a
 facility of the National Science Foundation operated under cooperative 
agreement by
 Associated Universities, Inc. This research was partially funded by NSF grants 
AST93-15285 and AST95-28394.

\newpage
\appendix

\section{Bayesian Pulsed Flux Estimation}

Pulsar flux densities are usually reported as phase-averaged quantities. 
Therefore, to
calculate a pulsed flux upper limit, one must 
assume a pulse width, pulse phase, off-pulse mean, and off-pulse rms.
To avoid trial and error estimates,
we have developed a method to calculate a PDF for the
pulsed flux density {\it independent} of these quantities.

We assume an $M$-bin folded pulse profile, where the amplitude of each bin is 
described
by $d_{i}$ ($i\in 1, M$). Our model for $\hat{d_{i}}$, the expected amplitude 
within
each bin, is parametrized by \nl
\indent \hspace{0.5in} {\it w} - width of square pulse \nl
\indent \hspace{0.5in} {\it m} - first bin of pulse \nl
\indent \hspace{0.5in} {\it F} - phase-averaged pulsed flux \nl
\indent \hspace{0.5in} $\mu$ - off-pulse mean \nl
\indent \hspace{0.5in} $\sigma$ - off-pulse rms

Given this model, we may express $\hat{d_{i}}$ as
\be
 \hat{d_{i}} = \left\{ \begin{array}{ll}
                FM/w + N_{i} & \hspace{0.3in} \mbox{\it i \rm = \it m, m+w$-$1} 
\\
                N_{i}   & \hspace{0.3in} \mbox{otherwise}
                \end{array}
        \right.
\label{eq:pulseshape}
\ee
where $N_{i}$ is the amplitude of the noise in bin $i$.
The likelihood function of the data is given by the joint probability density 
function
(PDF) of the noise samples.
Assuming Gaussian noise with mean $\mu$ and rms $\sigma$, and statistical 
independence of
the bins, we may write this as
\be
{\cal L} = (2\pi\sigma^{2})^{-M/2} \exp 
\left[-\frac{1}{2\sigma^{2}}\sum_{i=1}^{M}(N_{i}-\mu)^{2}\right].
\label{eq:like}
\ee

To find the best model parameters, we use Bayes' Theorem to write the posterior
probability $p(F,w,m,\mu,\sigma|{\bf d, I})$ of a model
parametrized by $F, w, m, \mu,$ \& $\sigma$ given the data {\bf d} and any prior 
information
{\bf I} as
\be
p(F,w,m,\mu,\sigma|{\bf d, I}) = \frac{p({\bf d}|F,w,m,\mu,\sigma,{\bf I}) 
p(F,w,m,\mu,\sigma|{\bf I})}{p({\bf d}|{\bf I})}.
\label{eq:bayes}
\ee
The factor $p({\bf d}|F,m,w,\mu,\sigma,{\bf I})$ is simply the likelihood 
function
given in Eq.~\ref{eq:like}. Because we expect all parameter values to be equally
likely
a priori, we
assume flat priors for $F, w, m,$ \& $\mu$ (i.e. $p(F|{\bf I})$ = constant) and 
a prior
 $p(\sigma|{\bf I}) \propto 1/\sigma$ (i.e. $p({\rm log}(\sigma)|{\bf I})$ = 
constant). We choose a prior
that is uniform in $\log \sigma$ to express our ignorance of the scale of the 
noise variation. 
Then, we can express the posterior probability as simply
\be
p(F,m,w,\mu,\sigma|{\bf d, I}) \propto \frac{{\cal L}}{\sigma}.
\label{eq:posterior}
\ee

Since we would like to calculate the PDF of the phase-averaged
 flux $F$ independent of the other parameters,
we must marginalize the posterior probability
over the ``nuisance parameters'' $w, m, \mu,$ \& $\sigma$.
%We write the cumulative distribution function (CDF) for $\Fbar$ as
%\be
%F(\Fbar) = P(\frac{uw}{M} < \Fbar) = \int dw \int_{0}^{\frac{M\Fbar}{w}} dF 
%\int dm \int d\
%mu \int d\sigma \hspace{0.1in}  \frac{{\cal L}}{\sigma}
%\label{eq:cumdist}
%\ee
We write the PDF for $F$ as
\be
f_{F}(F) \propto \int\frac{dw}{w} \int dm \int d\mu \int d\sigma
\hspace{0.1in} p(F,w,m,\mu,\sigma|{\bf d, I}).
\label{eq:pdf}
\ee

Substituting Eq.~\ref{eq:like} into Eq.~\ref{eq:pdf}, we find that the integrals
over $\mu$ and $\sigma$ may be done analytically. After some algebra, we find
\be
f_{F}(F) \propto \int\frac{dw}{w} \int dm \left(D_{2} + \frac{F^{2}M^{2}}{w} - 
\frac{2FMD_{p}}{w} - \frac{(FM-D_1)^{2}}{
M} \right)^{{{(1-M)}/{2}}},
\label{eq:finalbayes}
\ee
where constant factors have been eliminated and
\be
D_{1} = \sum_{i=1}^{M}d_{i}, \hspace{0.2in} D_{2} = \sum_{i=1}^{M}d_{i}^{2},  
\hspace{0.2
in} \& \hspace{0.2in}  D_{p} = \sum_{i=m}^{m+w-1}d_{i}.
\ee

In the case that the width and/or phase of the pulse is known, the
marginalization over $m$ and/or $w$ in Eq.~\ref{eq:finalbayes} can be removed to
further constrain the PDF for $F$.

To test our method, we have created many simulated profiles consisting of a 
pulse
superimposed on Gaussian noise,
and have confirmed that the PDF of Eq.~\ref{eq:finalbayes} does maximize at the 
correct
value of $F$. Some example simulation results for square wave pulses are shown 
in Figure
9. Figure 10 shows the results of applying our algorithm to profiles consisting 
only of
Gaussian random noise.

\section{Bayesian Flux Estimation of a Scintillating Source}

We assume that a source's intrinsic flux density is modulated by interstellar 
scintillations
by a factor $g$, so that the measured flux density is
\be
F = gS + N,
\label{eq:f2}
\ee
where $N$ is the additive noise and $S$ is the intrinsic flux density.
To calculate a PDF for $S$ we assume we have
 $m$ measurements of $F$ and $m$ separate estimates
of $N$, corresponding to making on and off-pulse flux measurements.
We assume that the noise PDF is Gaussian (a good assumption for radio 
observations with large
time-bandwidth product), and
that we have estimated $\mu$, the average value of the off-pulse mean, and 
$\sigma$,
the standard deviation of this quantity, for each of $m$ profiles. The PDF of 
the phase-averaged
noise, $f_{N}(N)$, is a Gaussian with mean $\mu$ and standard deviation 
$\sigma$, and
the PDF for an individual measurement of $F$ is
\be
f_{F}(F) = \int dN f_N(N) f_g\left ( \frac{F - N}{S} \right),
\label{eq:dFpdf}
\ee
where  $f_g$ is the PDF of $g$.
The likelihood function is a combination of all the data using PDF's of this
form.  So, for $m$ statistically independent observations, the likelihood 
function for $S$ is
\be
{\cal L}(S) =
\prod_{j=1}^{m} f_{F}(F_j)
\ee

We cannot use this scheme for all cases because generally we need to
use the joint PDF for the scintillation gain $g$ for the different measurements.
Because there is no closed expression for this, we consider only two
limiting cases.

\subsection{Case I: Statistically Independent ISS}

The first case we consider is when the interstellar scintillations are 
statistically
independent between measurements.
For this case, we may write
\be
{\cal L}_S(S) = \prod_{j=1}^{m}
\int dN f_N(N) f_g\left ( \frac{F_j - N}{S} \right).
\label{eq:likesi}
\ee
where $f_N(N)$ is a Gaussian described by $\mu_j$ and $\sigma_j$.

For $\dtd \ll T$
(i.e. the ISS time scale $\dtd$ is much less than the averaging time $T$)
and/or
$\dnud \ll B$
(the ISS bandwidth is much less than the receiver bandwidth),
the ISS is quenched and has a PDF that is related to a $\chi^2$ PDF
with the number of degrees of freedom given by $2\niss$, where
$\niss$ is the number of scintles averaged over.   In this case,
statistical independence between observations is a good assumption.

\subsection{Case II: Perfectly Correlated Scintillations}
The second case we consider is when
the time scale for interstellar scintillations is much longer than the
{\it total} data acquisition time (e.g. $mT$ for contiguous measurements). In 
this case,
the scintillation gain $g$ is identical for all measurements, and has
some unknown value with PDF described by Eq.~\ref{eq:fgexp}.
In this case, individual measurements
are statistically independent for the noise but completely identical with 
respect to $g$.
The likelihood function is therefore simply the product of the noise PDFs, and 
can be
expressed as
\be
{\cal L}_S(S) = \prod_{j=1}^{m} f_{F}(F_j) =
\prod_{j=1}^{m} f_{N}(F_j-S).
\ee
We can use this equation to get the likelihood function for
the product $S^{\prime} \equiv g S$ and then integrate over $f_g(g)$ to
get the likelihood for $S$:
\be
{\cal L}_S(S) &=& \int dg \, f_g(g) {\cal L}_{S^{\prime}}(gS) \\
\label{eq:likeid}
{\rm where} \hspace{0.25in}
{\cal L}_{S^{\prime}}(S^{\prime}) &=&
     \prod_{j=1}^{m} f_{N}(F_j-S^{\prime}).
\ee

Once we have chosen the scintillation regime of interest, and calculated ${\cal 
L}_S(S)$ with
Eq.~\ref{eq:likesi} or Eq.~\ref{eq:likeid}, we may
calculate the PDF of $S$ as the normalized likelihood function
\be
f_S(S) = \frac  { {\cal L}_S(S)}
                { \int_0^{\infty} dS\, {\cal L}_S(S) }.
\label{eq:likenorm}
\ee

When no source contributes to the `on-source' measurement, we expect
$f_S(S)$ to maximize at $S=0$ and have a width determined by
$\sigma$ and the number of measurements.  An upper bound on $S$ can be
calculated by choosing a probability level $1-\epsilon$ and calculating
the value of $S$ such that the area of $f_S(S)$ above that value is
$\epsilon$.
If $f_S(S)$ maximizes at a non-zero value, then a confidence interval for $S$ 
can be
similarly calculated.

When, as in the calculation of an upper limit using the method of Appendix A,
we do not have a single measured value for $F_{j}$, but
instead a PDF, we must replace $F_{j}$ by an integral over its PDF. In this
case, Eq.~\ref{eq:likesi} becomes
\be
{\cal L}_S(S) = \prod_{j=1}^{m} \int f_F(F_j) dF_j
\int dN f_N(N) f_g\left ( \frac{F_j - N}{S} \right).
\label{eq:totalint}
\ee

{}

\begin{figure}[h]
\centerline{\psfig{figure=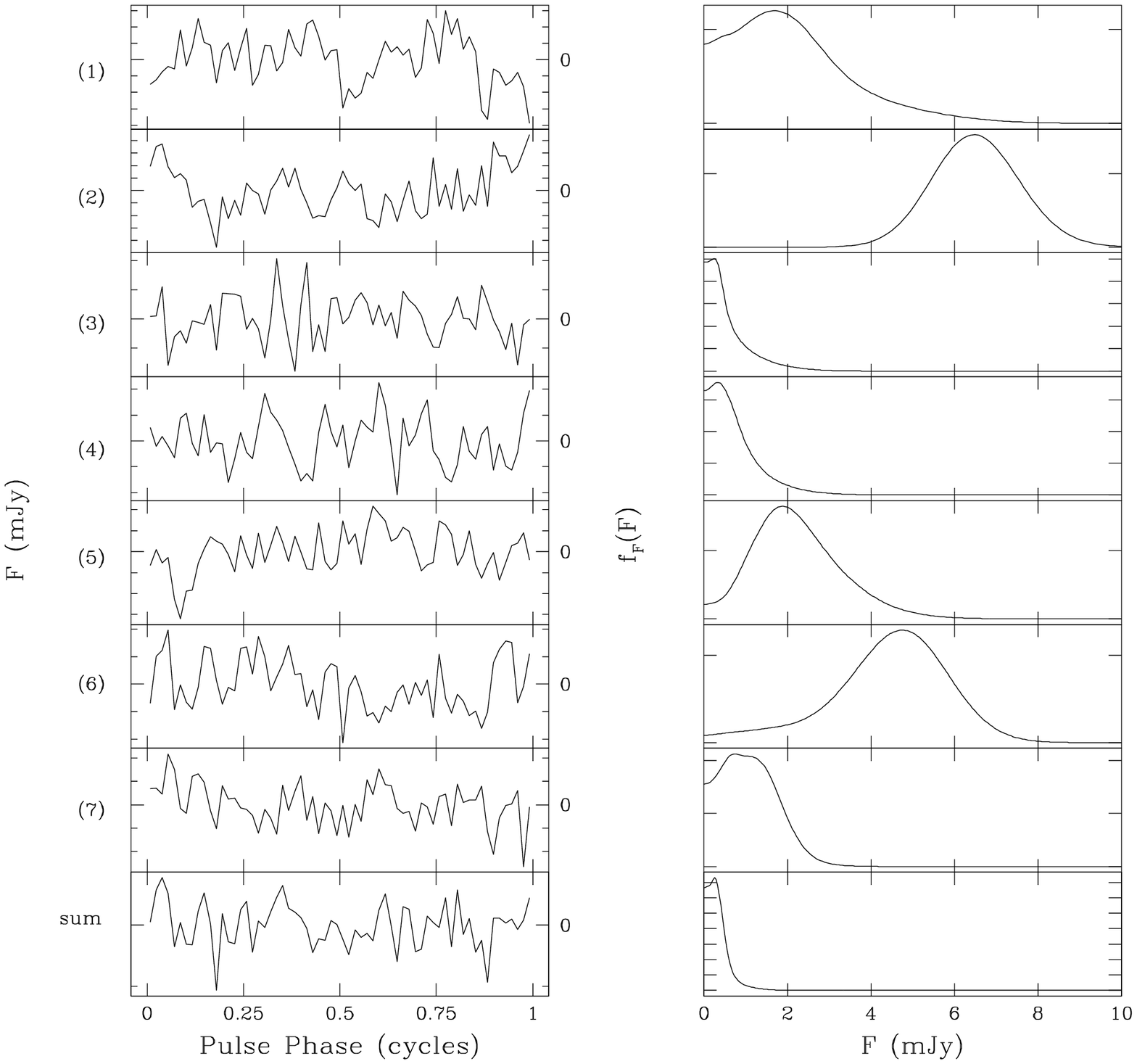,height=7.0in,width=6.0in}}
\caption{Left: The upper 7 plots show the 64 bin folded pulse profiles for 
the individual
data sets. The y-axis has units of mJys (one tick mark = 10 mJy) and have been
normalized to zero mean. The profiles have been shifted so
that the first bin of each profile corresponds to the same pulse phase. The 
bottom
plot shows the profiles from all 7 individual phase-referenced scans summed.
Right: The upper 7 plots show $f_{F}(F)$ for the individual Geminga scans.
The lower plot shows $f_{F}(F)$ for the summed profile.}
\end{figure}

\begin{figure}[h]
\centerline{\psfig{figure=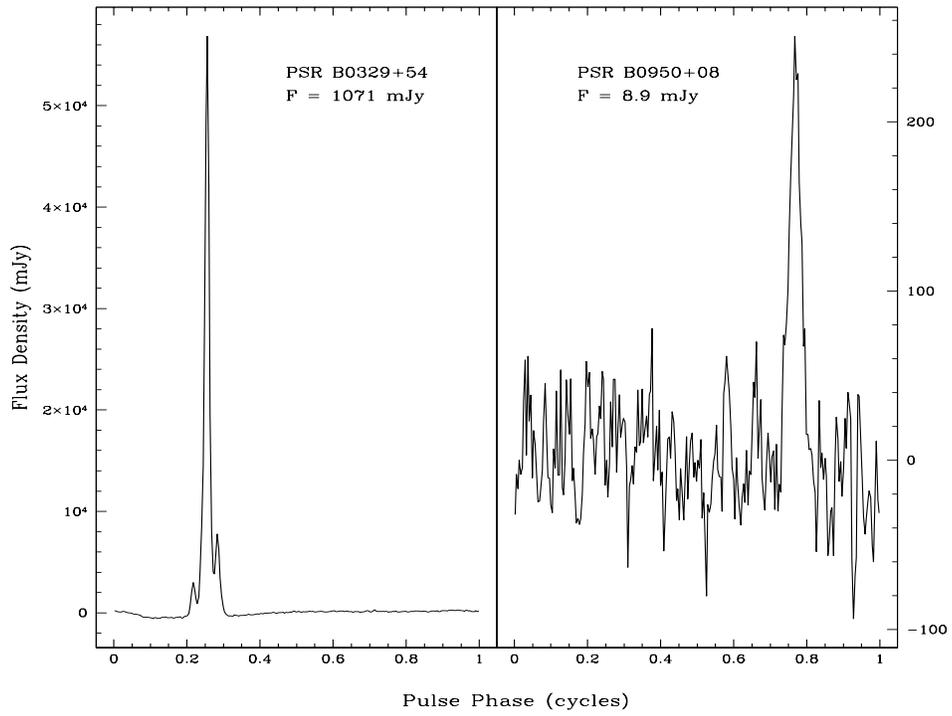,height=4.0in,width=5.0in}}
\caption{Folded pulse profiles for PSR B0329+54 (left) and PSR B0950+08 
(right)
are shown. The off-pulse mean has been subtracted.
The phase-averaged pulsed flux density is 1071 mJy for B0329+54
and 8.9 mJy for B0950+08.}
\end{figure}

\begin{figure}[h]
\centerline{\psfig{figure=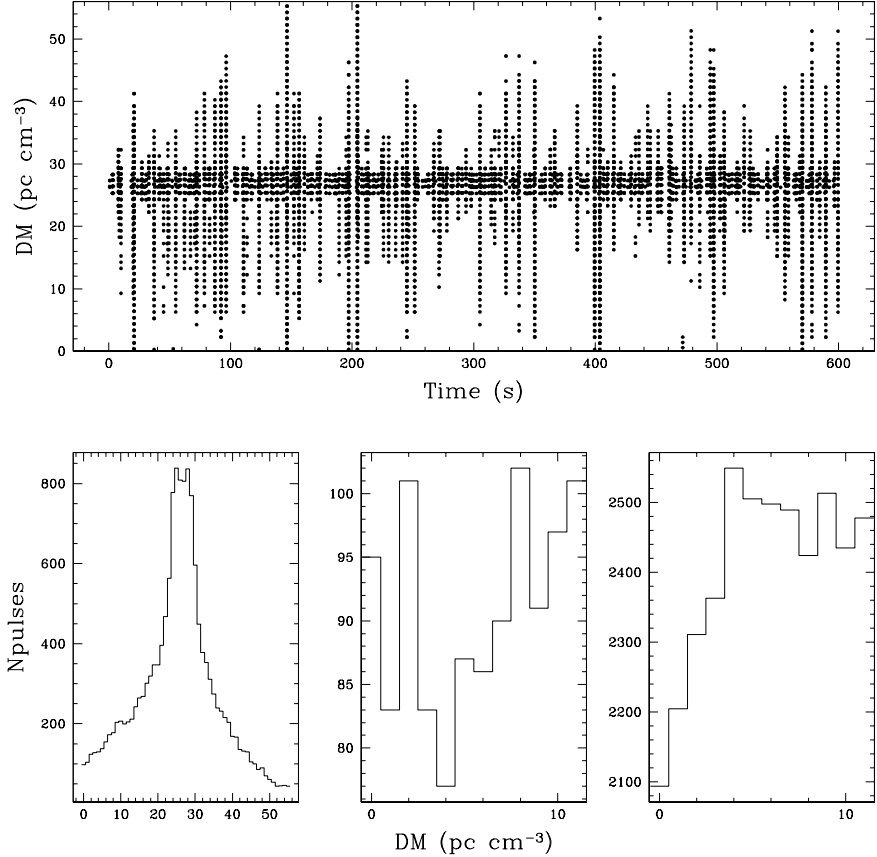,height=5.0in,width=5.0in}}
\caption{Upper: For PSR B0329+54, all individual pulses with amplitudes above 
9$\sigma$
have been plotted vs. DM and arrival time. This pulsar, at a
DM of 26.8 pc cm$^{-3}$, is clearly detectable through its individual pulses. 
Lower:
The number of isolated pulses above a  4 $\sigma$ threshold is plotted vs. DM for
(from left to right) B0329+54, B0950+08, and Geminga. While a peak at the
DM of PSR B0329+54 is again obvious, there is no evidence of individual pulses 
from PSR
B0950+08 or Geminga, both with DM $\approx$ 3 pc cm$^{-3}$. The individual 
pulses detected in these
directions are most likely due to the considerable interference at 317 MHz. We 
note that the
baseline of the histograms is not zero, and the distributions of pulses for PSR 
B0950+08
and Geminga are consistent with a uniform distribution in DM.
We also note that, while the number of pulses above threshold is 25 times
greater for Geminga than for B0950+08, the total time of the Geminga 
observations is greater
by a factor of 40, and so the {\it rate} of pulse detection is {\it lower} for 
Geminga.}
\end{figure}

\begin{figure}[h]
\centerline{\psfig{figure=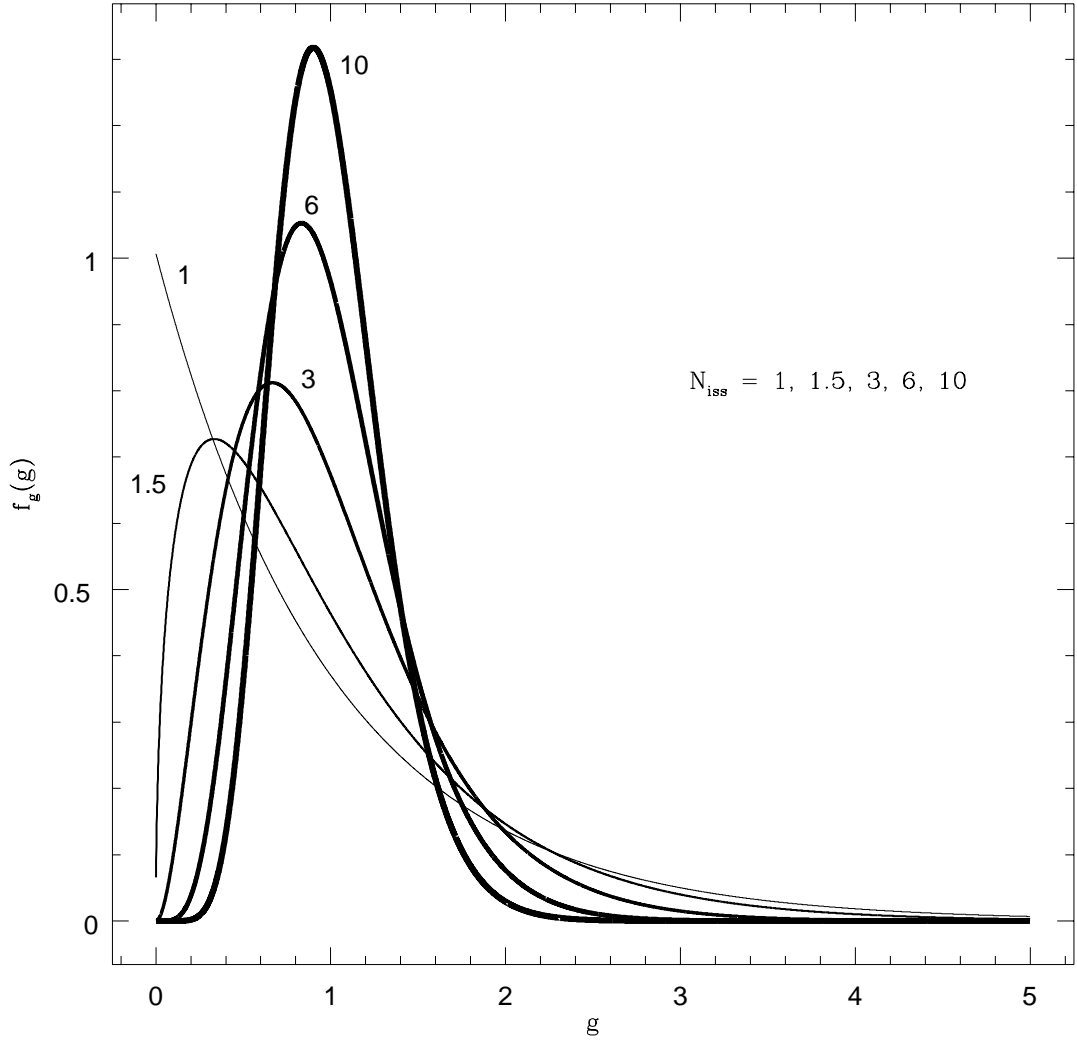,height=6.0in,width=6.0in}}
\caption{The PDF of g is plotted for different values of $\niss$, defined in
Eq. 6. The lines shown are for (from thinnest to thickest)
$\niss = 1, 1.5, 3, 6, 10$, and are tagged with their $\niss$ values.}
\end{figure}

\begin{figure}[h]
\centerline{\psfig{figure=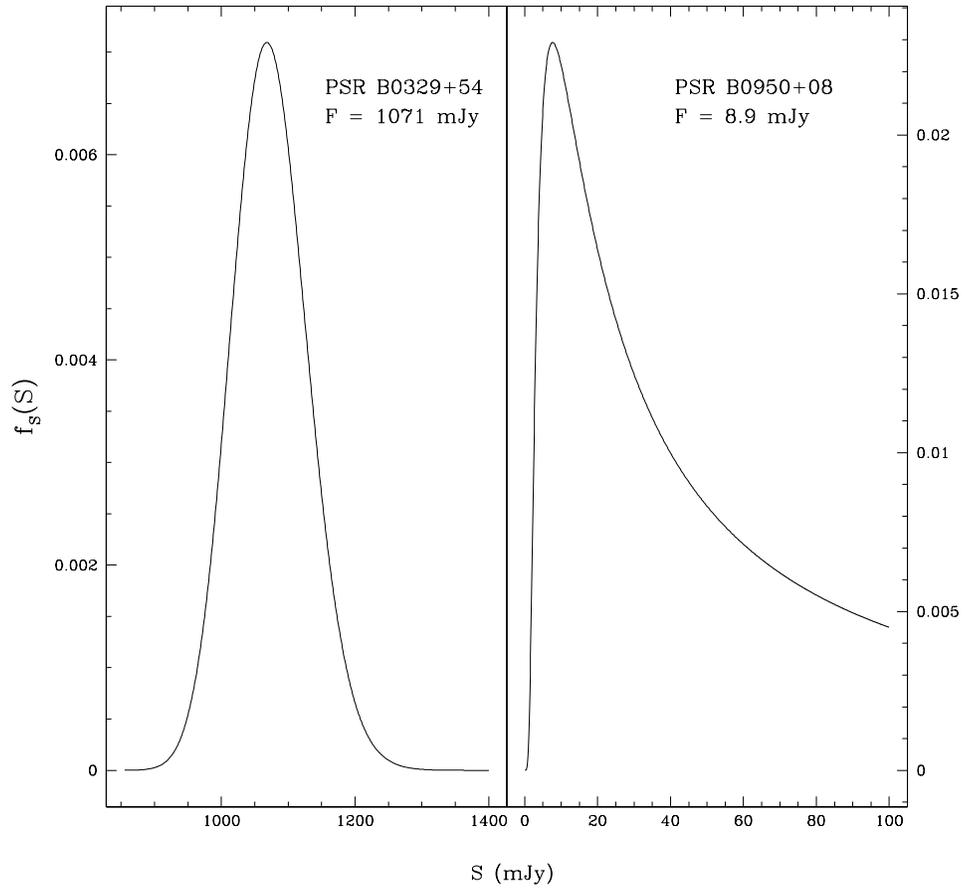,height=5.0in,width=5.0in}}
\caption{The PDFs for phase-averaged intrinsic source flux density $S$, 
given the measured
phase-averaged fluxes (from Figure 2), for our two test pulsars are shown. For 
PSR 
B0329+54, $\niss = 371$, and for PSR B0950+08, $\niss = 1.095$.}
\end{figure}

\begin{figure}[h]
\centerline{\psfig{figure=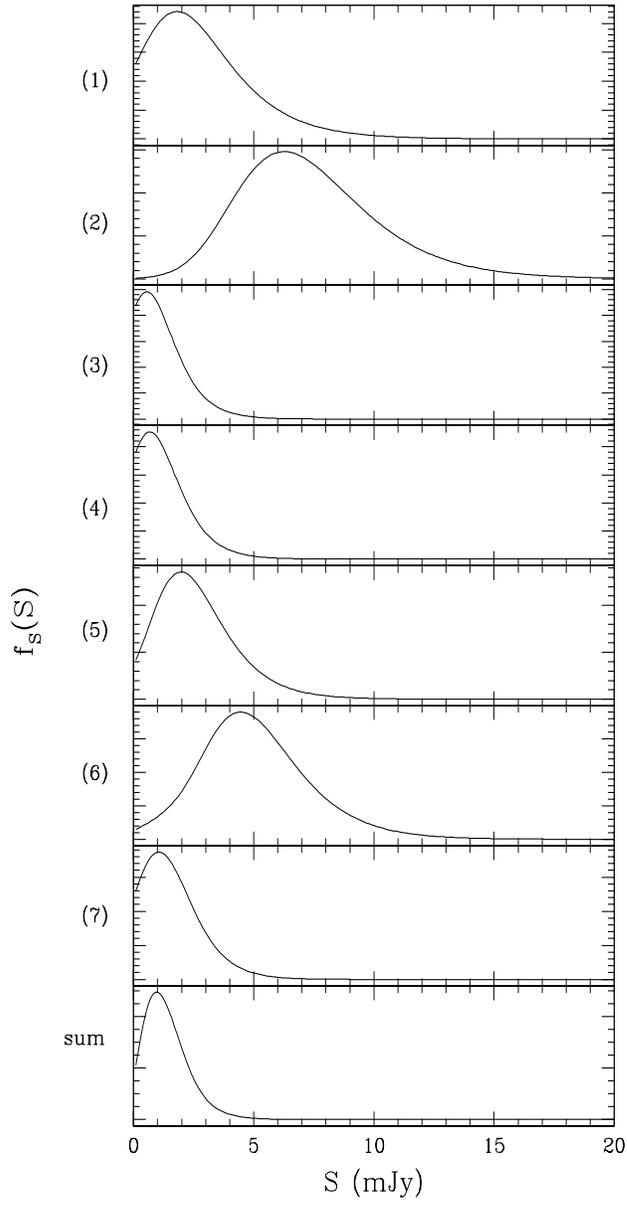,height=7.0in,width=6.0in}}
\figcaption{The upper 7 plots show ${f_S(S)}$ for the individual Geminga scans.
The lower plot shows ${f_S(S)}$ for the summed profile.}
\end{figure}

\begin{figure}[h]
\centerline{\psfig{figure=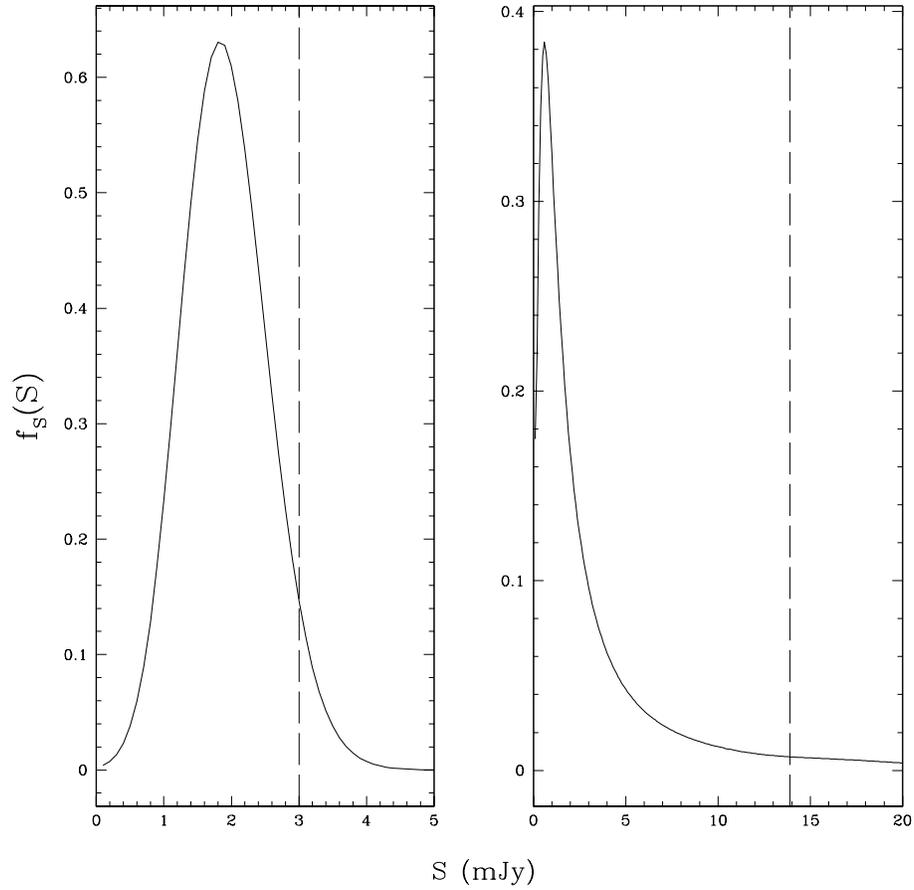,height=5.0in,width=5.0in}}
\caption{The left  plot shows the composite PDF for source flux density $S$ 
calculated
from all 7 Geminga scans, assuming statistical independence among measurements.
The right plot shows the composite PDF for $S$ calculated
in the limiting case where the timescale for ISS is {\it longer}
than the total time of our observation. The dotted lines mark the 95\%
confidence upper limits.}
\end{figure}

\begin{figure}[h]
\centerline{\psfig{figure=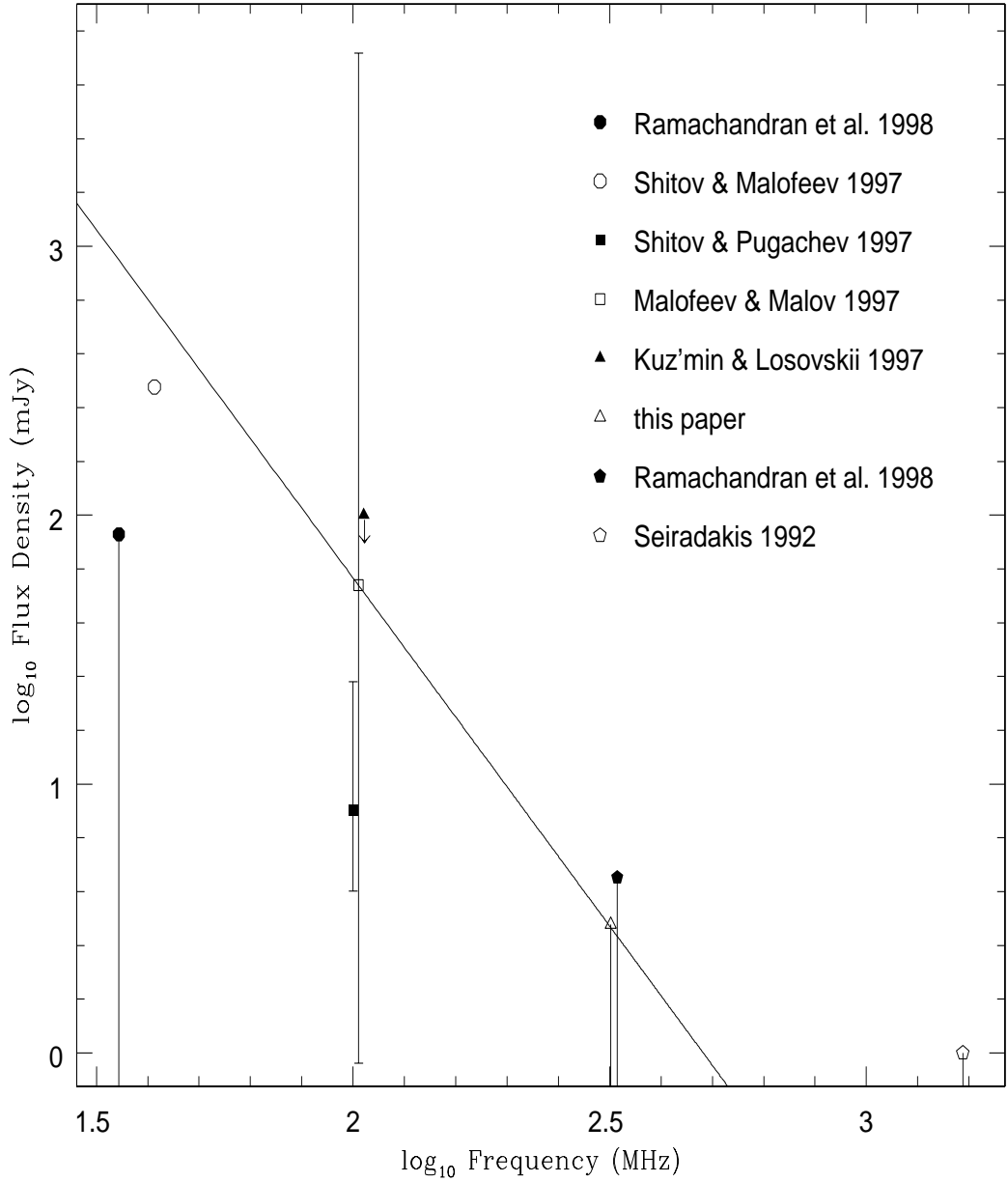,height=7.0in,width=6.0in}}
\caption{This plot shows all published Geminga upper limits and detections, 
along with
error bars (when available). The three 102.5 MHz detections are offset slightly from each other 
for clarity.
The slope of the solid line (2.65) represents our calculated lower limit on 
Geminga's
spectral index.}
\end{figure}

\begin{figure}[h]
\centerline{\psfig{figure=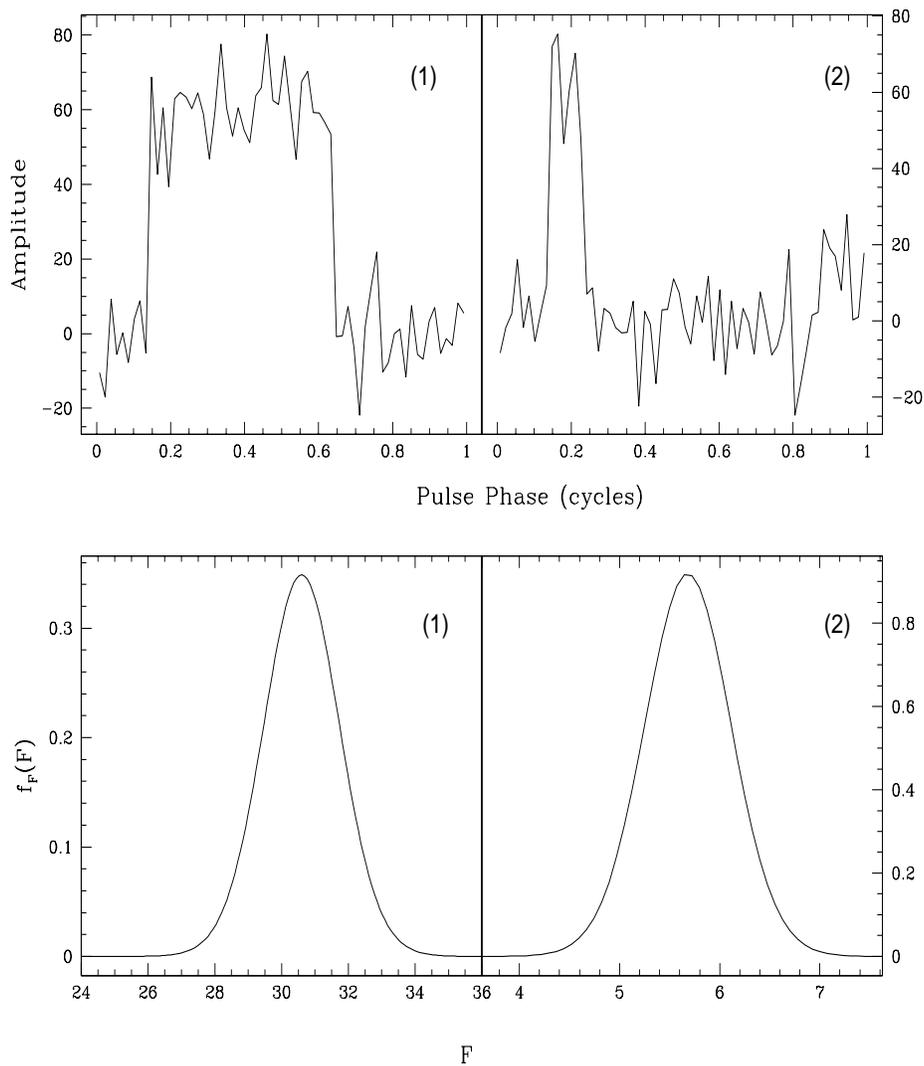,height=6.0in,width=5.0in}}
\caption{The top plots show two simulated pulse profiles with M = 64, $\mu = 
0$, and
$\sigma = 10$. For profile (1), $w = 32$ and $F = 30$, and for profile (2), $w = 
6$ and $F = 
5.625$.
Applying the Bayesian pulse detection scheme to these profiles results in
the PDFs shown in the bottom plots. Note that these PDFs
maximize at $F$, with scatter determined by the Gaussian noise in the original
profiles.}
\end{figure}

\begin{figure}[h]
\centerline{\psfig{figure=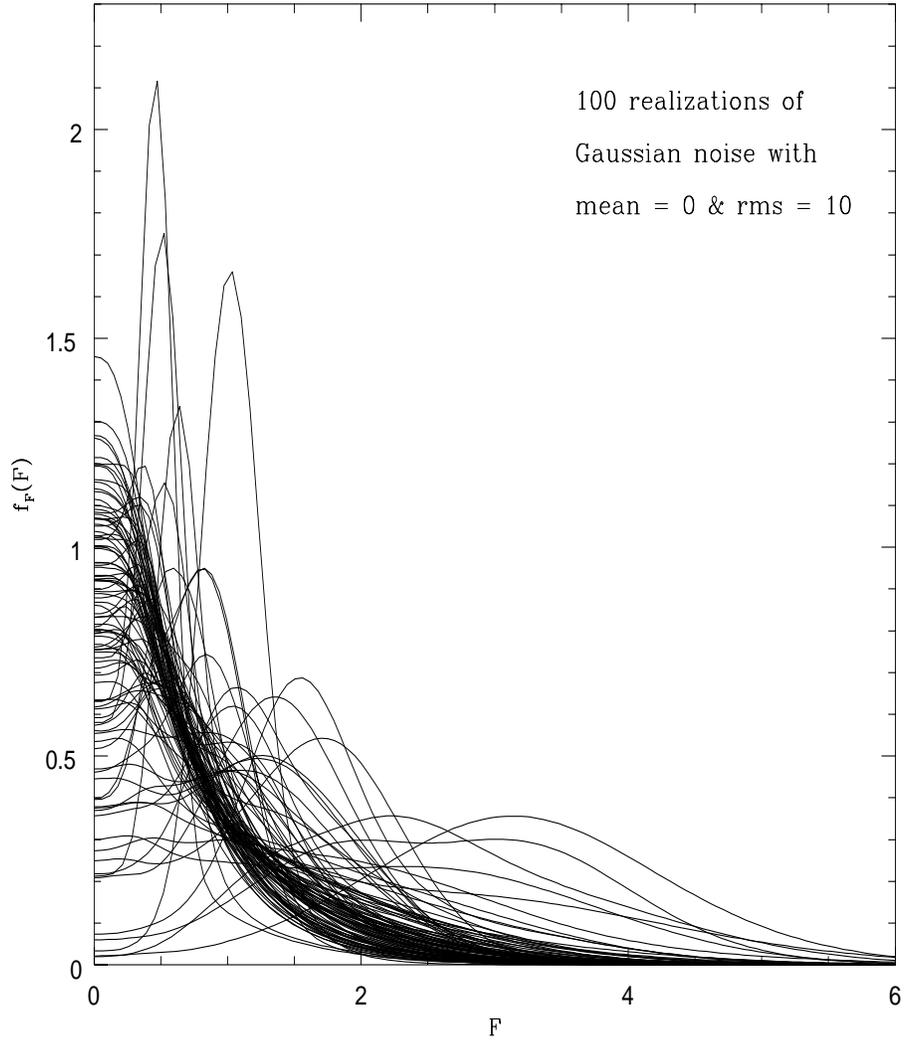,height=6.0in,width=5.0in}}
\caption{This plot show the PDFs resulting from application of the Bayesian 
pulse
detection scheme to 64 bin profiles of {\it only} Gaussian noise with
$\mu = 0$ and $\sigma = 10$. The results of 100 independent trials are plotted. 
Some 
PDFs maximize at non-zero values, all less than the rms noise level.}
\end{figure}
 
\end{document}